\pdfoutput=1
\documentclass[11pt]{article}
\usepackage{amssymb}
\usepackage{graphicx}
\parskip \baselineskip
\newcommand{\su}{\subseteq}

\newcommand{\ol}{\overline}

\newcommand{\ra}{\rightarrow}
\newcommand{\Ra}{\Rightarrow}

\begin{document}

\title{Compression with wildcards: All spanning trees with prescribed vertex-degrees}

\author{Marcel Wild}

\maketitle

{\it ABSTRACT. } We compactly encode (thus not one-by-one) all spanning trees of any graph $G$. This is achieved by processing all minimal cutsets of $G$. Surprisingly a 1986 algorithm of Winter seems to achieve (Conjecture 2)  exactly the same amount of compression, although in totally unrelated ways. Although Winter's algorithm is faster, our algorithm adapts to several variations of the problem in ways that are foreclosed to its competitor. One variation is the compact enumeration of all spanning trees $T$ that obey degree-conditions on the vertices $v$ of $G$; thus e.g. exactly 3 edges of $T$ must be incident with vertex $v_1$, exactly 1 edge with vertex $v_2$, and so forth.

\section{Introduction}

According to Kirchhoff's matrix-tree theorem (see e.g. Wikipedia) the number of spanning trees of a connected graph $G$ can be obtained in lightning speed by evaluating a certain determinant closely related to the adjacency matrix of $G$. We will use this (and say "by Kirchhoff") for control purposes because our main task is the  enumeration (=generation) of all these trees.
Starting with a crucial recurrence relation discovered in 1902 by Feussner (according to Knuth [K,p.462]), many algorithms have been proposed for that purpose. Twelfe of them are reviewed and pitted against each other in [CCCMP]. 

Here is how the author got involved in this matter. Since several years I strive to enumerate combinatorial structures (or models of Boolean functions) in a compressed format, using suitable types of wildcards. A general type of combinatorial structure is a transversal $X$ with respect to a hypergraph (=set system) $\cal H$, i.e. $X\cap H\neq\emptyset$ for all hyperedges $H\in {\cal H}$. The (transversal) $e$-algorithm of [W1] compresses the set $TR({\cal H})$ of all transversals. The present article  focuses on ${\cal H}= MCUT(G)$, where $ MCUT(G)$ is the set of all minimal cutsets of a graph $G$. In this case $TR({\cal H})$ is the set $ CONN(G)$ of all connected edge-sets of $G$. Having $ CONN(G)$ in compressed format immediately yields the reliability polynomial of $G$, but our main target is the subset $ST(G)\su CONN(G)$ of all spanning trees of $G$. As is well known, the spanning trees are exactly the $(n-1)$-element edge-sets among $CONN(G)$, where $n$ is the number of vertices of $G$. Pleasantly,  the compressed enumeration of $CONN(G)$ carries over to $ST(G)$.
 The overall procedure will be called {\it Mcuts-To-SpTrees}.
 
  This was already part of the preliminary\footnote{Let us mention, once and for all, a topic that {\it only} occurs in [W3], and not in the present article. It is a dual approach towards $ST(G)$ that is based on the cycles of $G$ instead of its minimal cutsets. As discussed in [W3], the possibility to tackle the problem in dual ways extends to the level of matroids.} version [W3] of the present article.
In order to trim [W3] for publication the author decided to pit Mcuts-To-SpTrees against the glorious winner of the contest in [CCCMP], i.e. against {\it Winter's algorithm} [Wi] of 1986.
That led to two surprises. First, and initially dissapointing, Winter's algorithm is considerably faster than Mcuts-To-SpTrees. Second, Winter's algorithm also achieves compression, but in  ways totally unrelated to minimal cutsets and wildcards. All the more surprising is it that there is a unifying data structure (called 01g-row) that allows to measure the compression of both,
and that this compression is exactly the  same!  The author has no idea why this is so.

Here comes the Section break-up. Section 2 describes Winter's algorithm on a toy graph $G_1$ that will accompany us throughout the article. Section 3 reviews basic facts about minimal cutsets and gives details about the $e$-algorithm from [W1,W2]. 

The core Section 4 formulates two conjectures and numerically supports them. The first concerns Mcuts-To-SpTrees on its own and needs some prerequisites to be stated. Upon running the $e$-algorithm on any hypergraph $\cal H$ it holds that $Min(TR({\cal H}))\su RMin(TR({\cal H}))$. Here $Min(TR({\cal H}))$ is the family of all (inclusion-)minimal ${\cal H}$-transversals and $RMin(TR({\cal H}))$ is the subfamily of certain {\it row-minimal} transversals $X\in TR({\cal H})$ which is immediate to retrieve  from $TR({\cal H})$. Conjecture 1 claims that for the very particular case ${\cal H}:= MCUT(G)$ 
it holds that $Min(TR({\cal H}))= RMin(TR({\cal H}))$, thus  $\su$ becomes $=$. Conjecture 2 relates Mcuts-To-SpTrees to Winter's algorithm and concerns the equal number of 01g-rows mentioned above.
When running Mcuts-To-SpTrees the intermediate dataset $CONN(G)\ (=TR({\cal H}))$ is interesting on its own; it yields at once the so called reliability polynomial of $G$. 

 Section 5 handles the case that $G$ is the complete graph $Comp(n)$ on $n$ vertices. Numerical evidence  suggests (Conjecture 3) that Conjecture 1 can then be refined in amazing ways. Suffice it to say that if Conjecture 3 holds, then an average 01g-row in the compressed representation of $ST(Comp(n))$ holds asymptotically more than $\frac{e^n}{n^2}$ spanning trees. Moreover, for small $n_0$  the stored representation of $ST(Comp(n_0))$ (we call this the "Library Method") shortcuts immensily (beating Winter's algorithm) the calculation of $ST(G)$ for {\it all} $(n_0,m)$-graphs $G$. 
 
 There are a few opportunities for Mcuts-To-SpTrees to make up for trailing Winter's algorithm (in case $n$ is too large for the Library Method). Namely, in Section 6 we adapt  Mcuts-To-SpTrees to the enumeration of relevant {\it parts} of ${\cal A}\su ST(G)$. Most attention is devoted to the set $\cal A$  of all $T\in ST(G)$ that satisfy prescribed degree conditions (as described in the Abstract). This naturally leads to edge-covers $X$ of $G$, i.e. edge-sets $X$ that are incident with all vertices. Section 7 ponders whether $ST(G)$ can also be sieved from the set $ECOV(G)$ of all edge-covers rather than from $CONN(G)$.

\section{Winter's algorithm}

 {\bf 2.1} It is best to describe Winter's algorithm on a toy example. Thus the vertices of the undirected graph $G_1$ in Figure 1 are labelled from $1$ to $n=5$. The $m=7$ edges need to be labelled as well but e.g. {\it not} as $\{2,5\}$ for the edge connecting vertices $2$ and $5$. This is because the {\it same} edge may connect {\it different} vertices throughout the algorithm. We chose (but this is by no means necessary) to label the edges $1$ to $m$ according to some obvious "lexicographic" way in which they {\it originally} connect vertices. There will be no danger  confusing the vertex labels $1,..,n$ and the edge labels $1,...,m$. We start by contracting the highest vertex $5$ into its highest neighbour, which happens to be $4$. As a result (see $G_2$) the smaller label survives, parallel\footnote{Instead of printing parallel edges we content ourselves  listing their labels.} edges may arise ($5,6$ respectively $3,4$), and the edge(s) lost by contraction must be remembered (see $\{7\}$). Instead of contracting edge $7$ we could have deleted it (triggering some graph $G_3$) but we persue that later (2.1). For the time being we stick to contraction and hence contract the highest vertex of $G_2$, i.e. $4$, into its highest neighbour, i.e. $3$. This results in $G_4$; note that the {\it accompanying information} $\{7\}$ has become $\{7,\{5,6\}\}$. Likewise $G_4$ gives rise to $G_5$, and $G_5$ triggers an edge-less 1-vertex graph $G_7$ with  accompanying information $\{\{7\},\{5,6\},\{3,4\},\{1,2\}\}$.  It signifies\footnote{For a proof we refer to Winter's article [Wi].} that each of the eight transversals of this set system  (such as $\{7,6,3,2\}$ or $\{7,5,4,1\}$, visualized in Fig.1 (e),(f)) is the edge set of a spanning tree of $G_1$.

\includegraphics[scale=0.55]{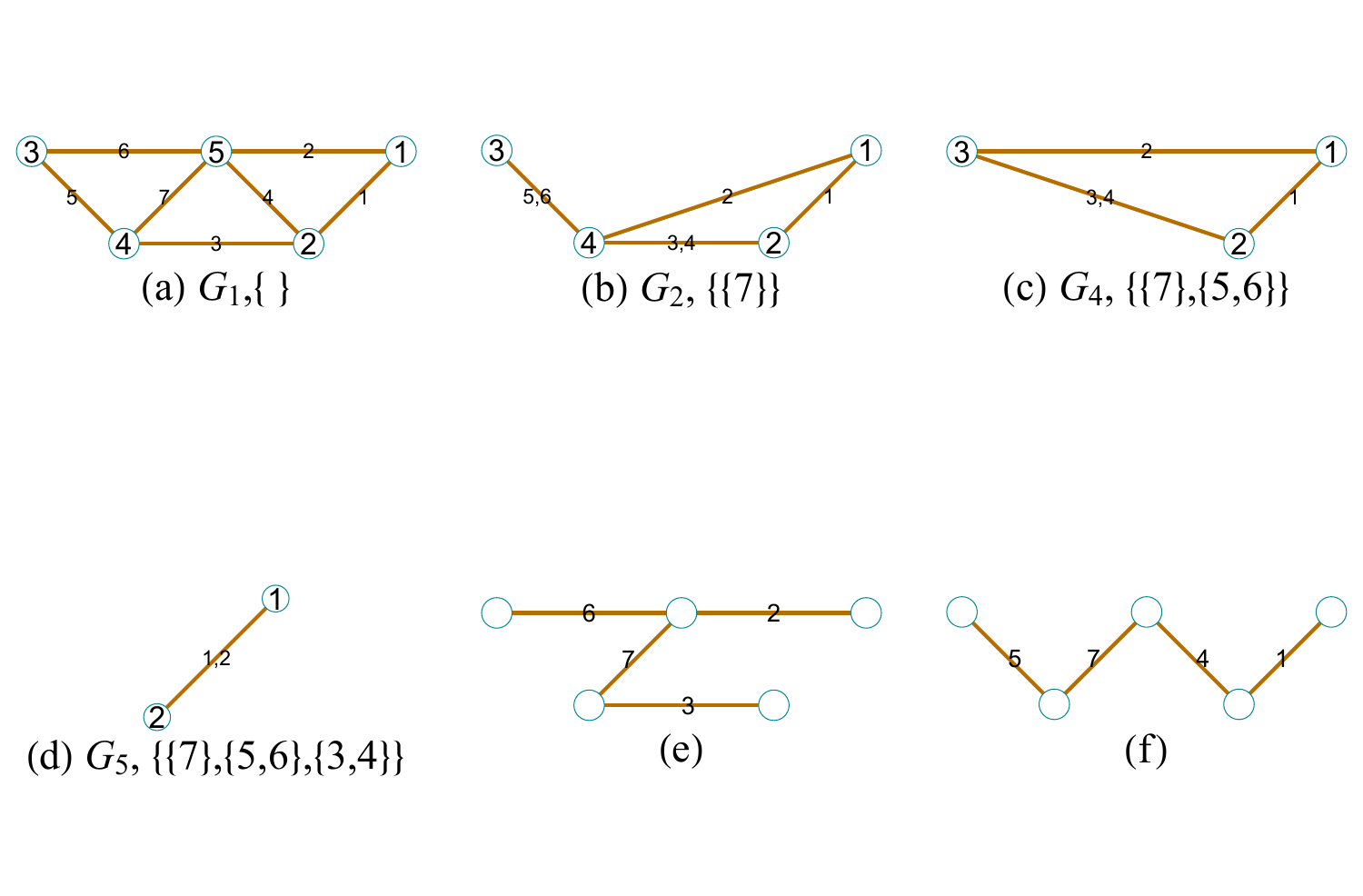}

{\sl Figure 1: The workings of Winter's algorithm.}

{\bf 2.2} As previously mentioned, instead of contracting the multiple edges between certain pairs of vertices, these multiple edges can also be deleted, unless this disconnects\footnote{Let us quote from the  bottom of page 276 in [CCCMP], i.e. from the description of Winter's algorithm: {\it Deletion of $S(n_i,n_j)$ occurs only when $n_i$ has adjacent vertices other than $n_j$.} The word 'only' is misleading; actually, one cannot avoid a full-blown connectivity test of the resulting graph. Nevertheless Winter's algorithm outperforms  its competitors in [CCCMP].} the graph. This gives rise to the computation tree in Figure 2. The labels on an arc from father $G_i$ to son $G_j$ are the selected edges of $G_i$ that were {\it  contracted} in order to get $G_j$.
If there is a second, unlabelled, arc from $G_i$ to $G_{j+1}$, then {\it deleting} the selected edges does not disconnect $G_i$ and thus yields a second son $G_{j+1}$ of $G_i$. 

\includegraphics[scale=0.85]{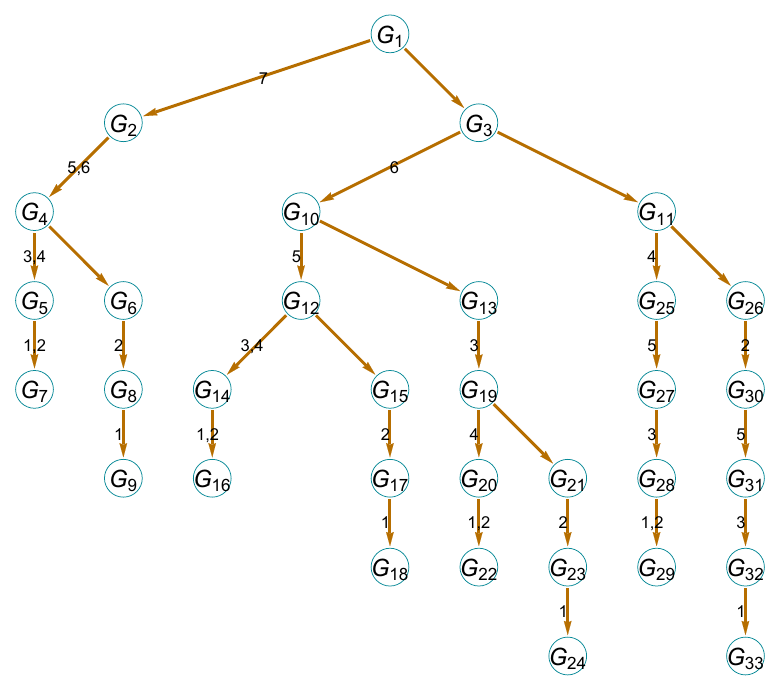}

{\sl Figure 2: This computation tree arises when Winter's algorithm is applied to $G_1$.}

The computation tree in Figure 2 has root $G_1$ and eight leaves $G_7,G_9,..,G_{33}$. The path from $G_1$ to $G_7$ has been discussed in detail above. Generally the accompanying information of a leaf $G_i$ is obtained by collecting the edge-labels encountered on the path from $G_1$ to $G_i$.

 Let us introduce\footnote{This tool was introduced by the author previously, and in other contexts.} the crisper notation (say) $(g,g,1,1,0,1,0)$ for the accompanying information $\{\{6\},\{3\},\{4\},\{1,2\}\}$ of $G_{22}$. Generally a {\it g-wildcard} consists of at least two symbols $g$. If it occurs in a {\it 01g-row} like $(g,0,g,g,1)$ then it demands "exactly one bit 1 in the area of the $g$'s". Thus formally

$(1)\quad (g,0,g,g,1):=\{({\bf 1},0,{\bf 0},{\bf 0},1),({\bf 0},0,{\bf 1},{\bf 0},1),({\bf 0},0,{\bf 0},{\bf 1},1)  \}.$

 If several $g$-wildcards occur within the same 01g-row, they are distinguished by subscripts. Consequently $(g_1,g_1,g_2,g_2,g_3,g_3,1)$ contains exactly the  characteristic bitstrings of the eight spanning trees encoded by  $\{\{7\},\{5,6\},\{3,4\},\{1,2\}\}$. For later reference, here are the eight 01g-rows that match the accompanying information of the leaves of the computation tree in Figure 2:

\begin{tabular}{c||c|c|c|c|c|c|c|| c }
 & 1 & 2 & 3 & 4 & 5 & 6 &7  &  \\ 
&  &  &  &  &  &  &  & \\ \hline
$G_7\ra$ &   $g_1$ &   $g_1$ &   $g_2$   & $g_2$ & $g_3$ &  $g_3$ & $1$ & 8 \\ \hline
$G_9\ra$ &   $1$ &   $1$ &    $0$ &   $0$ &   $g$ &  $g$ & $1$ &2  \\ \hline
$G_{16}\ra$ &   $g_1$ &   $g_1$ &   $g_2$   & $g_2$ &   $1$ &  $1$ & $0$ &4  \\ \hline
$G_{18}\ra$ &   $1$ &   $1$ &   $0$ &   $0$ &   $1$ & $1$ &  $0$ & 1 \\ \hline
$G_{22}\ra$ &   $g$ &   $g$ &   $1$ &     $1$ &   $0$ & $1$ &  $0$ &2  \\ \hline
$G_{24}\ra$ &   $1$ &   $1$ &   $1$ &    $0$ &   $0$ & $1$ &  $0$ &1 \\ \hline
$G_{29}\ra$ &   $g$ &   $g$ &   $1$ &   $1$ &  $1$ & $0$ &   $0$  &2\\ \hline
$G_{33}\ra$  &   $1$   & $1$ &   $1$ &   $0$ & $1$ & $0$ & $0$  &1\\ \hline
\end{tabular}

{\sl Table 1: The compression of $ST(G_1)$ provided by Winter's algorithm}

The order $G_1,G_2,...,G_{33}$ in Figure 2 matches a depth first search (which gives preference to the left sons) of the computation tree. As is well known, depth first searches are best implemented by last-in-first-out (LIFO) stacks. To refresh memory let us, assuming the LIFO-stack opens to the right,  list its contents throughout the algorithm until it's empty. It  initially contains $G_1$, then its sons $G_3,G_2$ (abbreviated as $(1),(3,2)$, etc):

$ (1),(3,2),(3,4),(3,6,5),(3,6,7),(3,6),(3,8),(3,9),(3),(11,10),(11,13,12),(11,13,15,14),$

$(11,13,15,16),(11,13,15),(11,13,17),(11,13,18),(11,13),(11,19),(11,21,20),(11,21,22),$

$(11,21),(11,23),(11,24),(11),(26,25),(26,27),(26,28),(26,29),(26),(30),(31),(32),(33),(\ )$

 We emphasize that the 01g-rows derived from the output of Winter's algorithm are always mutually disjoint\footnote{Of course, it is argued in [Wi] as well that each spanning tree is only generated once.}. For instance, look at the third and fifth row in Table 1. The earliest predecessor of the corresponding leaves $G_{16}$ and $G_{22}$ is found to be $G_{10}$. One of its outgoing arcs has label 5, the other not. Accordingly all spanning trees in the third row contain edge 5, and all spanning trees in the fifth row do not contain edge 5.

\section{Viewing spanning trees as minimal transversals}

After some   terminology (3.1) we glimpse on the $e$-algorithm of [W1] which renders all transversals of a set system in a compressed format (3.2). Often only the minimal transversals are of interest. This leads us to spanning trees of $G$ because they can be viewed as the minimal transversals of the set system of all mincuts of $G$; this is dealt with in 3.3. and 3.4.

{\bf 3.1} For a fixed set $W$  we code subsets $X\su W$ as bitstrings as done already  in Section 2.  One often uses don't-care symbols like $*$ to indicate that both $0$ or $1$ are allowed at a specified position. We adopt this practise except that we write '2' (by obvious reason) instead of '$*$'. This leads to {\it 012-rows} like\footnote{We  gloss over the fact that $(1,0,1)$, viewed as 012-row (without 2's) , strictly speaking is $\{(1,0,1)\}$.}

$(2)\quad r=(0,2,0,0,2,1):=\{  (0,{\bf 0},0,0,{\bf 0},1),\   (0,{\bf 0},0,0,{\bf 1},1),\  (0,{\bf 1},0,0,{\bf 0},1),\  (0,{\bf 1},0,0,{\bf 1},1)  \}.$

Thus, as a set system $r= \{  \{6\},\ \{6,5\},\ \{6,2\},\ \{6,2,5\}  \}$, which sometimes will be trimmed to $ \{6,65,62,625\}$ or $\{6,56,26,256\}$. The following notation is self-explanatory:

$(3)\quad zeros(r):=\{1,3,4\},\ ones(r):=\{6\},\ twos(r):=\{2,5\}.$

 More inventive than replacing $*$ by $2$ is the {\it e-wildcard} $(e,e,\cdots, e)$ which means 'at least one $1$ here'. Like for $g$-wildcards, distinct $e$-wildcards are distinguished by subscripts and are wholly independent of each other.  Instead of a formal definition of the arising {\it 012e-rows} (which can be found in [W1]), two examples may suffice:

$({\bf e_1},0,e_2,0,{\bf e_1},0,e_2)=\{{\bf 1}3,17,137,\ {\bf 5}3,57,537,\ {\bf 15}3,157,1537\}$

$r_0:=(1,0,2,1,2,2,\ e_1,e_1,\ e_2,e_2,\ e_3,e_3,e_3,e_3)\Ra |r_0|=\ 2^3\cdot (2^2-1)^2\cdot (2^4-1)=1080.$

{\bf 3.1.1}  Let $Min({\cal S})$ be the family of all (inclusion-)minimal members of a set system ${\cal S}$. If ${\cal S}=r$ for some 012e-row $r$, then calculating $Min(r)$ is easy and all sets in $Min(r)$ are equicardinal. Roughly put, set all 2's to 0 and choose exactly one 1 in each $e$-wildcard, hence turning it into an $g$-wildcard. For instance

$\hspace*{5cm}Min(r_0)=(1,0,0,1,0,0,g_1,g_1\ ,g_2,g_2,\ g_3,g_3,g_3,g_3).$

Generally suppose the 012e-row $r'$ has $t$ many $e$-wildcards of lengths $\epsilon_1,...,\epsilon_t$. Putting 

$\hspace*{5cm} deg(r'):=|ones(r')|+t,$

 it holds that

\begin{itemize}
\item[(4)] $|Min(r')|=\epsilon_1\epsilon_2\cdots\epsilon_t$, and $|X|=deg(r')$ for all $X\in Min(r')$.
\end{itemize}

{\bf 3.2} The introduced  012e-rows arise naturally as intermediate and final data structure in the following procedure. Given any hypergraph (=set system) ${\cal H}=\{H_1,\ldots,H_h\}\su {\cal P}(W)$, an $\cal H$-{\it transversal} is any set $X$ such that   $X\cap H_i\not = \emptyset$ for all $i$ in $[h]:=\{1,2,...,h\}$.
The  {\it (transversal) e-algorithm} of [W1] calculates the family ${TR}({\cal H})$ of all ${\cal H}$-transversals as a {\it disjoint} union of 012e-rows, thus ${TR}({\cal H})=\sigma_1\uplus\cdots\uplus \sigma_t$. 
As in Winter's algorithm the computation tree is handled with a LIFO-stack, but the LIFO-stack now contains 012e-rows and the (potentially more than two) {\it candidate sons} $r'$ of the top LIFO row $r$ are calculated in wholly different ways. A candidate son $r'$ is {\it feasible} (i.e. survives and becomes a 'proper' {\it son}) iff it contains an $\cal H$-transversal. Fortunately this can be decided at once since it is easy to see [W1,p.126] that

\begin{itemize}
\item[(5)] {\it $r'$ is infeasible iff  $H_i\su zeros(r')$ for some $i\in [h]$.}
\end{itemize}

By (5),  and because each LIFO top row $r$ is either {\it final} (=leaf of the computation tree) or has at least one son, the $e$-algorithm runs [W1,p.127] in output-polynomial time $O(Nh^2w^2)$, where $w=|W|$ and $N$ is the number of transversals.

{\bf 3.2.1} It is an obvious yet crucial fact that each $X\in Min({TR({\cal H})})$ is row-minimal within the row $\sigma_i$ in which it happens to occur. Conversely, $\sigma_i$-minimal sets need not be minimal. Hence 

 $(6)\quad Min({TR}({\cal H}))=Min(\sigma_1\uplus\cdots\uplus \sigma_t)    \su Min(\sigma_1)\uplus\cdots\uplus Min(\sigma_t).$ 

We henceforth write

 $(7)\quad RMin({TR}({\cal H})):=Min(\sigma_1\uplus\cdots\uplus \sigma_t)$ 

for the family of all row-minimal sets.
Although not reflected in notation, $RMin({TR}({\cal H}))$ also depends on the {\it order} in which $\cal H$ is fed to the e-algorithm.

{\bf 3.2.2} In [W2] we strove to sieve $Min(TR({\cal H}))\cap r$ from $r$ for each $r$. In view of (6) the following terminology turned out to be handy. Call $r$ {\it bad} if $Min(r)$ contains no minimal $\cal H$-transversal. Otherwise $r$ is {\it good}, and it is {\it  very-good} if $Min(r)\su Min(TR({\cal H}))$. In particular,  each {\it minimum} (cardinality) transversal occurs in a  very-good row.

{\bf 3.3} In order to establish the connection to spanning trees in 3.4 we first recall some well known facts about mincuts. Let $G=(V,E)$  be a connected graph with vertex set $V=\{v_1,\ldots,v_n\}$ and edge set $E$. A {\it cutset} is a set $H\su E$ of edges whose removal results in a disconnected graph $(V,E\setminus H)$.
A {\it mincut}\footnote{Be aware that some authors define a 'mincut' as a cutset of minimum cardinality. They constitute a subset of our kind of mincuts.} is a minimal cutset. In this case $(V,E\setminus H)$  has exactly {\it two} connected components. The converse is true as well: Suppose $V=V_1\uplus V_2$ is a {\it good} partition of $V$ in the sense that $V_1,\ V_2$ are nonvoid and the induced subgraphs $G.V_1$ and $G.V_2$ are both connected. Then the set $H$ of all edges between $V_1$ and $V_2$ is a mincut.  (It follows that $G$ can have at most $\frac{2^n-2}{2}=2^{n-1}-1$ mincuts.)
We will sometimes identify a mincut of $H\su E$ with that {\it v}ertex-{\it s}et $V_i$ of its accompanying good partition $V=V_1\uplus V_2$  that has $1\in V_i$. The so arising {\it vs-mincuts} of $G_1$ are

$(8)\quad   \{1\},\{1,2\},\{1,5\},\{1,2,4\},\{1,2,5\},\{1,3,5\},\{1,2,3,4\},\{1,2,3,5\},\{1,2,4,5\},\{1,3,4,5\}.$

The matching (ordinary) mincuts, in shorthand notation, are these:

$(9)\quad MCUT(G_1)=\{12,234,1467,2457,367,1457,2467,357,56,134\}.$

\includegraphics[scale=0.65]{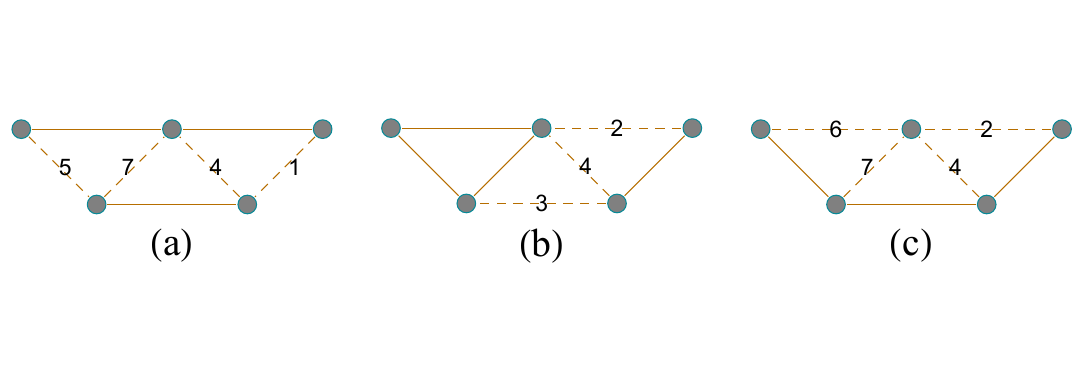}

{\sl Figure 3: Some of the mincuts listed in (9)}

The vs-mincuts in (8) are in {\it canonical order} in the sense that after $\{1\}$, all cardinality 2 vs-mincuts, then all cardinality 3 vs-mincuts, and so forth, are listed. Within vs-mincuts of the same cardinality we order lexicographically (in the obvious sense). In the numerical experiments of Section 4 we generate the set $MCUT(G)$ of all mincuts of $G$ by testing all $2^{n-1}-1$ partitions of $V$, and selecting the good ones. There are more clever ways to get $MCUT(G)$ (see [TSOA]) but for small $n$ our brute-force (and easier to implement) method suffices.

{\bf 3.4} Generally, let $G=(V,E)$ be connected, and let ${\cal H}=MCUT(G)$. As is well known and easy to see, then ${TR}({\cal H})$ is the family $CONN(G)$ of all connected\footnote{In the sense that $(V,X)$ is a connected graph in the usual sense. In particular, $(V,X)$ has no isolated vertices.} edge-sets $X$ of $G$.  For instance, feeding   $MCUT(G_1)$ (displayed in (8))  to the $e$-algorithm yields 
 $CONN(G_1)$  in Table 2:

\begin{tabular}{c||c|c|c|c|c|c|c||c  }
 & 1 & 2 & 3 & 4 & 5 & 6 &7&    \\ 
&  &  &  &  &  &  &  & \\ \hline
$r_1=$ & $1$ & $0$ & $1$ & $0$ & $e$ & $e$ & $1$&3  \\ \hline
$r_2=$ & $1$ & $0$ & $0$ & $1$ & $1$ & $1$ & $0$&1  \\ \hline
$r_3=$ & $1$ & $0$ & $e_1$ & $1$ & $e_2$ & $e_2$ & $e_1$&9  \\ \hline
$r_4=$ & $e$ & $1$ & $1$ & $e$ & $1$ & $0$ & $0$&3  \\ \hline
$r_5=$ & $e$ & $1$ & $1$ & $e$ & $0$ & $1$ & $0$&3  \\ \hline
$r_6=$ & $1$ & $0$ & $1$ & $0$ & $1$ & $1$ & $0$&1  \\ \hline
$r_7=$ & $e$ & $1$ & $e$ & $e$ & $1$ & $1$ & $0$&7  \\ \hline
$r_8=$  & $e_1$ & $1$ & $e_1$ & $e_1$ & $e_2$ & $e_2$ & $1$&21  \\ \hline
\end{tabular}

{\sl Table 2: Compact representation of $ CONN(G_1)={TR}( MCUT(G_1))$ as $r_1\uplus\cdots\uplus r_8$.}

Adding up the rightmost column in Table 2 shows that $G_1$ has $3+1+\cdots+21=48$ connected edge-sets.
The 012e-rows in Table 2 are actually 01e-rows, and so

$(10)\quad RMin({TR}( MCUT(G_1))=\ol{r}_1\uplus\cdots\uplus \ol{r}_8,$

where $\ol{r}_i$ is obtained by merely\footnote{Recall from 3.1.1 that generally also the $2$'s must become $0$'s. Whether feeding hypergraphs of type ${\cal H}= MCUT(G)$ always yields 01e-rows, remains an open problem.}  replacing each $e$-wildcard with the matching $g$-wildcard:

\begin{tabular}{c||c|c|c|c|c|c|c||c  }
 & 1 & 2 & 3 & 4 & 5 & 6 &7&    \\ 
&  &  &  &  &  &  &  & \\ \hline
$\ol{r_1}=$ & $1$ & $0$ & $1$ & $0$ & $g$ & $g$ & $1$&2  \\ \hline
$\ol{r_2}=$ & $1$ & $0$ & $0$ & $1$ & $1$ & $1$ & $0$&1  \\ \hline
$\ol{r_3}=$ & $1$ & $0$ & $g_1$ & $1$ & $g_2$ & $g_2$ & $g_1$&4  \\ \hline
$\ol{r_4}=$ & $g$ & $1$ & $1$ & $g$ & $1$ & $0$ & $0$&2  \\ \hline
$\ol{r_5}=$ & $g$ & $1$ & $1$ & $g$ & $0$ & $1$ & $0$&2  \\ \hline
$\ol{r_6}=$ & $1$ & $0$ & $1$ & $0$ & $1$ & $1$ & $0$&1  \\ \hline
$\ol{r_7}=$ & $g$ & $1$ & $g$ & $g$ & $1$ & $1$ & $0$&3  \\ \hline
$\ol{r_8}=$  & $g_1$ & $1$ & $g_1$ & $g_1$ & $g_2$ & $g_2$ & $1$&6  \\ \hline
\end{tabular}

{\sl Table 3: The expected superset $RMin({TR}( MCUT(G_1))$   of $ST(G_1)$ actually coincides with it.}

\section{The first two Conjectures}

Recall that the minimal connected edge-sets of any connected graph $G$ are exactly its spanning trees, and also coincide with the $(n-1)$-element connected edge-sets.  From (6) and (7) follows:

$(11)\quad ST(G)=Min( CONN(G))=Min({TR}( MCUT(G))\su  RMin({TR}( MCUT(G)).$

In particular $ST(G_1)\su  RMin({TR}( MCUT(G_1))$.
From  $|ST(G_1)|=21$ (by Kirchhoff or by Table 1) and $| RMin({TR}( MCUT(G_1))|=2+1\cdots +6=21$ (Table 3) follows 
$ST(G_1)=  RMin({TR}( MCUT(G_1))$. Both Table 1 and Table 3 have eight 01g-rows. Interestingly, the  number of 01g-rows {\it always} seems to be the same\footnote{ However, the actual 01g-rows in Table 3 do not coincide with the ones in Table 1, not even upon permuting the coordinates. For instance $\ol{r}_8$ features $(g_1,g_1,g_1)$, whereas all g-wildcards in Table 1 have length at most two. }.  Specifically we make two conjectures:

\begin{enumerate}
\item[] {\bf Conjecture 1:} For each connected graph $G$
not only  $ST(G)\su RMin({TR}( MCUT(G))$ holds (fact (11)), but actually  $ST(G)= RMin({TR}( MCUT(G))$.
\item[] {\bf Conjecture 2:} Let $G$ be connected and suppose $ MCUT(G)$ is fed to the $e$-algorithm in the order that matches the canonical order of the vs-mincuts. Then the number of $01g$-rows produced by the $e$-algorithm equals the number  of $01g$-rows produced by Winter's algorithm.
\end{enumerate}

We henceforth call {\it Mcuts-To-SpTrees} the particular application of the $e$-algorithm described in Conjecture 2.
 Table 4 below supports Conjectures 1 and 2 numerically. Specifically,  $T_{Win}$ and $T_{McST}$ are the CPU-times (in seconds) of Winter's algorithm and Mcuts-To-SpTrees respectively. Both algorithms were implemented with Mathematica\footnote{It comes in handy that Mathematica has a hardwired command to test connectivity of graphs; cf the footnote in 2.2. Also Kirchhoff's  determinant is hardwired.}. Furthermore $R_{Win},R_{can},R_{ran}$ are the numbers of output 01g-rows produced by, respectively, Winter's algorithm, Mcuts-To-SpTrees with (vs-)mincuts processed in canonical order, and Mcuts-To-SpTrees  with mincuts processed in random order.
It turns out that Winter's algorithm is faster than Mcuts-To-SpTrees; the proportional gap even seems to grow with the number of mincuts to be imposed. Using Kirchhoff  the number of spanning trees $|ST(G)|$ could always be calculated in a fraction of a second, and it always coincided with $| RMin({TR}(MCUT(G))|$ in accordance with Conjecture 1. Furthermore always $R_{Win}(G)=R_{can}(G)$, in accordance with Conjecture 2. Notice that $R_{ran}$ coincided with $R_{can}$ for the three random (15,20)-graphs and  the three random (20,30)-graphs, yet differed slightly from $R_{can}$ in the other cases.

\begin{tabular}{|c|c|c|c|c|c|c|c|  }
  (n,m) & $|ST(G)|$ &mincuts & $T_{Win}$ & $T_{McST}$ & $R_{Win}$ & $R_{can}$  &  $R_{ran}$    \\ 
  &  &  &  &  &  & & \\ \hline
 (10,20) & 22266 &224& 2s & 6s & 2088 &2088    &  \\ \hline
  (10,20) &  &&  & 6s &  &    &2092  \\ \hline\hline

 (10,30) & 1'565'116 &421& 37s & 110s & 33156 & 33156 &  \\ \hline
 (10,30) &  & &  & 180s &  &    & 33216 \\ \hline
 (10,30) &  & &  & 186s &  &    & 33199 \\ \hline\hline

 (11,50) & 853'863'120 &1017& 1782s & 16223s & 2'143'440 & 2'143'440   &  \\ \hline\hline

 (15,20) & 2640 &120& 2s & 3s & 726 & 726   &  \\ \hline
 (15,20) &       & &    & 3s &      &    & 726 \\ \hline
 (15,20) &        & &     & 3s &     &    & 726 \\ \hline\hline

 (15,25) & 164'296 &1068& 42s & 311s & 21056 & 21056   &  \\ \hline
              &              &         &       & 408s &            &    &21057  \\ \hline
             &               &        &        & 366s &            &   & 21107 \\ \hline\hline
 (15,30) & 4'715'056 &3290& 400s & 10286s & 280'014 & 280'014   &  \\ \hline
 (15,30) &            &   &      & 10513s &         &    & 281'104 \\ \hline\hline

 (20,30) & 190'800 &460& 40s & 124s & 12864 & 12864   &  \\ \hline
 (20,30) &         &   &     & 133s &      &    &12864  \\ \hline
 (20,30) &         &   &      & 123s &      &    &12864  \\ \hline

\end{tabular}

{\sl Table 4: Numerical comparison of Winter's algorithm and Mcuts-To-SpTrees}

{\bf 4.1} To recap, in the present context the general hypergraph ${\cal H}$ from Section 3.2 equals ${\cal H}=MCUT(G)$. Then {\it all} mininimal ${\cal H}$-transversals are minimum (of cardinality $n-1$), and so each good row is  very-good. Rephrasing Conjecture 1 it claims that Mcuts-To-SpTrees produces no bad rows, and whence only  very-good rows. 

If Conjecture 1 is false, how much work is it to weed out the bad rows? Let $r$ be an arbitrary 012e-row. Recall from 3.1.1 that $deg(r)=p$ implies $|X|\ge p$ for all $X\in r$. Thus $deg(r)\le n-1$ is necessary for $r$ to contain (edge-sets of) spanning trees $T\su [m]$. Specifically:

\begin{itemize}
	\item[(12)] Let $G$ be a connected graph with $n$-element vertex set $V$ and (for convenience) edge-set $[m]$. Let $r$ be a $012e$-row of length $m$ with
	 $deg(r)\le n-1=|V|-1$. Then $r$ contains a spanning tree $T\su [m]$  iff
	\begin{itemize}
		\item[(i)] $[m]\setminus zeros(r)$ is a connected edge-set, and
		\item[(ii)] at least one edge-set $X\in Min(r)$ is independent (=cycle-free).
	\end{itemize}
	\end{itemize}

{\it Proof of (12).} Suppose $T\su [m]$ is a spanning tree of $G$ and $T\in r$. With $T$ also $[m]$ must be connected, and because of $deg(r)\le n-1=|T|$, there is $X\in Min(r)$ with $X\su T$. Evidently $X$ is independent. Conversely, suppose (i) and (ii) are satisfied. If $E:=[m]\setminus zeros(r)$ and $X\in r$ is arbitrary then by  definition of 012e-row for all $T$ with $X\su T\su E$ we have $T\in r$. In particular, the independent edge-set $X$ of  (ii) extends to a maximal independent edge-set $T$ of the (by (i)) {\it connected} graph $G'=(V,E)$, and so $T$ is a spanning tree of $G'$ (and thus $G$).
 By the above, $T\in r\ \square .$

Testing connectivity (i) is fast but testing (ii) becomes the more cumbersome\footnote{For any $t\ge 1$ and $\epsilon\ge 2$ it is easy to construct a graph $G$ and corresponding row $r$ consisting of $t$ many $e$-wildcards of length $\epsilon$ such that $r$ violates (ii) (albeit $ones(r)$ {\it is} independent). But for {\it random} rows $r$ it clearly holds that the larger $Min(r)$ the likelier (ii) holds.} the larger the $e$-wildcards and the higher their number $t$. For 012-rows $r$ condition (ii) boils down to checking the independency of only one set $ones(r)$.

{\bf 4.2} Given a connected graph $G$ we let $Rel(G,p)$ be the probability that $G$ stays connected when each edge gets deleted, independently from the others, with probability $1-p$ (i.e. stays 'operational' with probability $p$). It is easy to see [C,p.46] that $Rel(G,p)$ is a polynomial in $p$ which can be written as $Rel(G,p)=\sum_{i=0}^m c_i p^i(1-p)^{m-i}$ where $c_i$ is the number of connected\footnote{Let us check the extreme cases. Since $c_0=0$ (assuming $|V|\ge 2$) the right hand side evaluates for $p=0$ to $c_0\cdot 0^0\cdot 1^m+0+\cdots+0=c_0\cdot 1=0$, as is to be expected. Similarly, for $p=1$ the right hand side becomes $0+\cdots+0+c_m\cdot 1^m\cdot 0^0=1$, as it must.} edge-sets of cardinality $i$. Recall  that the main effort of Mcuts-To-SpTrees goes into representing $ CONN(G)$ as a disjoint union of 012e-rows. Since the number $\gamma_i(r)$ of $i$-element sets in the 012e-row $r$ is easily  calculated  [W1, Theorem 1], one obtains the coefficient $c_i$ of $Rel(G,p)$ by  summing up the numbers $\gamma_i(r)$ when $r$ ranges over the disjoint rows constituting $ CONN(G)$.

{\bf 4.2.1} Can Winter's algorithm be used to calculate the $c_i$'s as well? This is desirable in view of its speed and seems plausible since (a) also Winter's algorithm represents $ST(G)$ as disjoint union of 01g-rows, and  (b) $ CONN(G)$ consists of all supersets $Y\supseteq X$ obtained by letting $X$ range over $ST(G)$. To illustrate how  to collect all $Y$'s, take $G=G_1$ and imagine all $g$-wildcards in Table 1 have been replaced by corresponding $e$-wildcards. The originating 012e-rows constitute a subset of $CONN(G_1)$. Adding up their cardinalities yields $27+3+9+1+3+1+3+1=48$, which happens to be $|CONN(G_1)|$ (see 3.4). 

However, whether the 012e-rows obtained this way {\it generally exhaust} $CONN(G)$ remains an open question. 
If no, then the slower Mcuts-To-SpTrees must be invoked.

{\bf 4.2.2} Speaking of "slow", note however that every algorithm based on depth-first-search can be parallelized and thus, bluntly speaking, can be sped up to any desired degree.
This well-known fact is best argued within the equivalent LIFO framework. Namely, at any moment all 012e-rows in the (main) LIFO stack can be distributed to any number of satellite computers which launch their own LIFO stacks in order to process their received 012e-rows. The results are reported back to the main control.

\section{Complete graphs and the Library Method}

The present Section  focuses on the {\it complete}  $n$-vertex graph $G=Comp(n)$. By Cayley's Theorem $Comp(n)$ has exactly $n^{n-2}$ spanning trees. Furthermore, all
 $2^{n-1}-1$ proper partitions $V=V_1\uplus V_2$ are good, and so the $2^{n-1}-1$ mincuts of $Comp(n)$ are readily obtained. 

Let us investigate in detail what happens for $n=8$. Upon
 feeding the $2^{n-1}-1=127$ mincuts of $Comp(8)$ to the e-algorithm packs the $8^6=262144$ spanning trees into exactly $(n-1)!=5040$ many 012e-rows of length ${{n}\choose{2}}=28$. Furthermore, the  maximum number of row-minimals in a row is $(n-1)!$, which is e.g. achieved by the 01e-row $\ol{r}$ in Table 5 that features six $e$-wildcards. For ease of notation they are written as $AA,BBB$ up to $FFFFFF$, rather than $e_1e_1$ up to $e_6e_6e_6e_6e_6e_6$.

      \begin{tabular}{c|c|c|c|c|c|c|c|c|c| c|c|c|c|c|}
& 1 & 2 & 3& 4 & 5 & 6 &7 & 8 & 9  &10  &11  &12 &13  &14   \\ \hline 
&  &  & &  &  &  & &  &   &  &  &     &  &    \\ \hline
$\overline{r}=$ &  F & E & D & C & B & A & $1$ & F &  F & F & F & F &F &E     \\ \hline
 \end{tabular}

\begin{tabular}{c|c|c|c|c|c|c|c|c|c| c|c|c|c|c|}
& 15 & 16 & 17& 18 & 19 & 20 &21 & 22 & 23  &24  &25  &26 &27  &28   \\ \hline 
&  &  & &  &  &  & &  &   &  &  &     &  &    \\ \hline
 &  E & E & E & E & D & D & D & D &  C & C & C & B &B &A     \\ \hline
 \end{tabular}

{\sl Table 5: This 01e-row packs 5040 spanning trees of the complete graph $Comp(8)$  }

The edge labeling  is again the lexicographic one (2.1), so e.g. the last label $28$ goes to the edge connecting vertices 7 and 8. Because $|ones(\ol{r})|=1$ and $\ol{r}$ has six $e$-wildcards, all  $2\cdot 3\cdots 7=5040$ many $\ol{r}$-minimal sets have cardinality $7$ (see (4)). Because $7=|V|-1$,  all row-minimal sets are spanning trees. In other words, 
every
  transversal of $\{A,A\},\ \{B,B,B\},\ldots, \{F,F,F,F,F,F\}$ yields a spanning tree. Two random instances (a),(b) are shown  in Figure 4.

\includegraphics[scale=0.65]{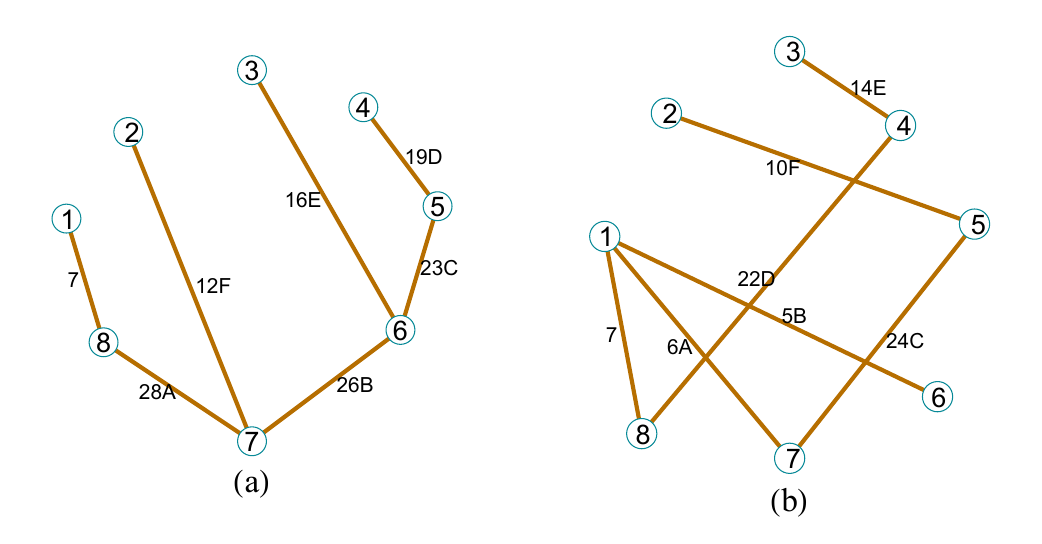}

{\sl Figure 4: Two out of 5040 spanning trees contained in row $\ol{r}$ of Table 5}

 Extrapolating the evidence from $n=8$ (and in fact all $n\le 11$) we put forth

{\bf Conjecture 3:}  {\it Feeding the $2^{n-1}-1$ many mincuts of $Comp(n)$ to the $e$-algorithm in canonical order yields exactly $(n-1)!$disjoint 012e-rows whose row-minimal sets match the spanning trees of $Comp(n)$. Moreover,  some of these rows contain exactly $(n-1)!\ $ spanning trees. }

 Provided  Conjecture 3 is true, the average number $av(n)$ of spanning trees contained in a 01e-row grows exponentially with $n$. Specifically, it follows from $n!\approx \sqrt{2\pi n}
\big(\frac{n}{e}\big)^n$ (Stirling's formula) that 

$$(12)\hspace*{3cm} av(n)=\frac{n^{n-2}}{(n-1)!} = \frac{n^{n-1}}{n!}\approx n^{n-1}\frac{1}{\sqrt{2\pi n}  } \bigg( \frac{e}{n}\bigg)^n >\frac{e^n}{n^2}.$$


{\bf 5.1} Sticking to $n=8$, consider the spanning graph $G'$ of $Comp(8)$  with edge set \\
 $E':=\{2,4,6,7,8,11,13,17,19,20,22,24,27\}$. Here 'spanning' means that each of the eight vertices is incident with some edge in $E'$. As a consequence each spanning tree $T$ of $Comp(8)$
which happens to satisfy $T\su E'$ is a spanning tree of $G'$ as well.
 Recall  that $Comp(8)=r_1\uplus\cdots\uplus r_{7!}$ for some disjoint 01e-rows $r_i$. By the above ${\cal ST}(G')=\biguplus_{i=1}^{5040}(r_i\cap r_{E'})$, where $r_{E'}$ is defined by $twos(r_{E'}):=E',\ zeros(r_{E'}):=[28]\setminus E'$. Of course $r_i\cap r_{E'}=\emptyset$ iff either $ones(r_i)\cap zeros(r_{E'})\neq\emptyset$ or some $e$-wildcard of $r_i$ is contained in $zeros(r_{E'})$. If $r_i$ is the row $\ol{r}$ of Table 5 then  $r_i\cap r_{E'}\neq\emptyset$. In fact

(13)\quad $r_1\cap r_{E'}=(0,E,0,C,0,1,1,F,0,0,F,0,F,0,0,0,E,0,D,D,0,D,0,C,0,0,1,0),$

and so the spanning trees of $G'$ contained in $r_1\cap r_{E'}$  match the 36 transversals of $CC,DDD,EE,FFF$.

{\bf 5.2}  It is clear that the arguments given for $G'$ above generalize to arbitrary $n$ and graphs $G=([n],E)$, provided $G$ is a {\it  spanning} subgraph of $Comp(n)$.
To fix ideas, let $CONN(Comp(n))=r_1\uplus\cdots \uplus r_N\ (N=(n-1)!)$, and let $r_E$ be defined as $r_{E'}$ above. Then ${\cal ST}(G)=\biguplus_{i=1}^{N}(r_i\cap r_{E})$.
 This tempts one to set up, for moderate $n$-values, a 'Library' of the $N$ many 01e-rows of length ${n}\choose{2}$ triggered by $Comp(n)$ (Conjecture 3).  This {\it Library Method} allows the compression of ${\cal ST}(G)$  without the need to compute, nor process all mincuts of $G$. We implemented the Library Method for $n=11$. Applying it to the (11,50)-graph of Table 4 took 52 sec, thus much faster than the 1782 sec of Winter's method.

 In order to efficiently catch the many Library rows $r_i$ with $r_i\cap r_E\neq\emptyset$ we employ a technique called {\it Vertical Layout} whose details are described in [W3].

\section{Enumerating all degree-restricted spanning trees}

While Winter's algorithm exhausts $ST(G)$ faster\footnote{The extra popularity that Winter's algorithm may achieve through the present article is well deserved!} than Mcuts-To-SpTrees and doesn't need the mincuts, the latter is better suited for enumerating certain types of spanning trees. This is because  Mcuts-To-SpTrees can be launched on {\it any} 012e-row; it need not be $(2,2,..,2)$. As to "certain types", we mainly dwell on degree-conditions, and glance at something else in 6.4.

{\bf 6.1} Suppose we only need to know the set $\cal D$ of those $T\in ST(G_1)$ that satisfy  $|T\cap\{3,5,7\}|=1$. It is clear that 

${\cal D}=ST(G_1)\cap(r_1\uplus r_2\uplus r_3)=(ST(G_1)\cap r_1)\uplus(ST(G_2)\cap r_2)\uplus(ST(G_1)\cap r_3),$

where $r_1,r_2,r_3$ are from Table 6:

\begin{tabular}{c|c|c|c|c|c|c|c|  }
 & 1 & 2 & 3 & 4 & 5 & 6 &7    \\ 
&  &  &  &  &  &  &   \\ \hline
$r_1=$ & $2$ & $2$ & ${\bf 1}$ & $2$ & ${\bf 0}$ & $2$ & ${\bf 0}$    \\ \hline
$r_1\cap ST(G_1)=$ & $0$ & $1$ & ${\bf 1}$ & $1$ & ${\bf 0}$ & $1$ & ${\bf 0}$    \\ \hline
$\uplus$ & $1$ & $e$ & ${\bf 1}$ & $e$ & ${\bf 0}$ & $1$ & ${\bf 0}$    \\ \hline\hline

$r_2=$ & $2$ & $2$ & ${\bf 0}$ & $2$ & ${\bf 1}$ & $2$ & ${\bf 0}$    \\ \hline
$r_2\cap ST(G_1)=$ & $1$ & $0$ & ${\bf 0}$ & $1$ & ${\bf 1}$ & $1$ & ${\bf 0}$    \\ \hline
$\uplus$ & $e$ & $1$ & ${\bf 0}$ & $e$ & ${\bf 1}$ & $1$ & ${\bf 0}$    \\ \hline \hline

$r_3=$ & $2$ & $2$ & ${\bf 0}$ & $2$ & ${\bf 0}$ & $2$ & ${\bf 1}$    \\ \hline
$r_3\cap ST(G_1)=$ & $1$ & $0$ & ${\bf 0}$ & $1$ & ${\bf 0}$ & $1$ & ${\bf 1}$    \\ \hline
$\uplus$ & $e$ & $1$ & ${\bf 0}$ & $e$ & ${\bf 0}$ & $1$ & ${\bf 1}$    \\ \hline

\end{tabular}

{\sl Table 6: Enumerating only certain $T\in ST(G_1)$}

The set system (say) $r_2\cap ST(G_1)$ is calculated by imposing the mincuts of $G_1$ upon $r_2$. The result will be a disjoint union of 012e-rows, in our case the 5th and 6th row in Table 6. Likewise $r_1,\ r_3$ are treated, and it follows that

 ${\cal D}=\{2346,1236,1346,1456,1256,2456,1467,1267,2467\}.$

  It appears that Winter's algorithm cannot  be adapted to this kind of task. If so, one needs to generate all of $ST(G)$ and then look in which 01g-rows the sought spanning trees are situated.  For $G=G_1$ the nine trees in $\cal D$ occupy the first six rows in Table 1. 
  
{\bf 6.2} The condition  $|T\cap\{3,5,7\}|=1$ above demands (see Figure 1(a)) that the degree of vertex 4 in $T$ be 1, i.e. $deg(T,4)=1$. What if degree conditions $deg(T,v_i)=d_i$ for {\it several} vertices $v_i$ must be obeyed? It is easy to translate this into a Boolean function $f$. Say $G=(V,[m])$ has $V=\{v_1,v_2,v_3,...\}$ with $star(v_1)=\{1,2,3\},
star(v_2)=\{3,4,5\},star(v_3)=\{1,5,6,7\}$, such that $deg(v_1)=1,deg(v_2)=deg(v_3)=2$ must be satisfied. Upon coupling a Boolean variable $x_i$ to each edge $i$, the Boolean function becomes

$f(x_1,...,x_7)=\Big((x_1\wedge\ol{x_2}\wedge\ol{x_3}) \vee (\ol{x_1}\wedge x_2\wedge\ol{x_3})\vee (\ol{x_1}\wedge\ol{x_2}\wedge x_3) \Big) \\
\hspace*{2cm}\wedge \Big((x_3\wedge x_4\wedge\ol{x_5}) \vee (x_3\wedge \ol{x_4}\wedge x_5)\vee (\ol{x_3}\wedge x_4\wedge x_5) \Big) \\
\hspace*{2cm}\wedge \Big((x_1\wedge x_5\wedge \ol{x_6}\wedge\ol{x_7}) \vee (x_1\wedge \ol{x_5}\wedge x_6\wedge \ol{x_7})
\vee\cdots \vee (\ol{x_1}\wedge \ol{x_5}\wedge x_6\wedge x_7) \Big)$

The models $y$ (=satisfying truth assignments) of $f$ match those edge-sets $Y\su [7]$ that satisfy all degree conditions; one example being $y=(0,0,1,0,1,0,1)$ and $Y=\{3,5,7\}$. There may be edges not involved in degree conditions, and accordingly we deal with 012-rows like $r=(0,0,1,0,1,0,1,2,2..,2)$. If $Y$ contains a circuit then $r$ contains no spanning trees. Otherwise applying Mcuts-To-SpTrees to $r$ generates all spanning trees contained in $r$ in
 polynomial total time. The occurence of even few 0's and 1's in $r$ speeds up Mcuts-To-SpTrees considerably, as testified in 6.3.

{\bf 6.2.1} Given a conjunction of conditions of type $deg(T,v_i)=d_i$, enumerating all edge-sets satisfying these conditions cannot be done in polynomial total time (because this problem specializes to enumerating all proper matchings of a graph). However in practical applications this is not an issue when only few vertex-degrees are prescribed; simply apply (say) the Mathematica command {\tt BooleanConvert} to the triggered Boolean function $f$. This yields the starter 012-rows fast enough. We mention in passing that polynomial total time can be restored when switching from type $=d_i$ conditions to type $\ge d_i$. This will be proven in a future article discussed in Section 7. By other reasons, polynomial total time also takes place in 6.4.

{\bf 6.3} As to numerics, we picked the (11,50)-graph $G$ in Table 4 (and so $|T|=10$ for all 853'863'120 spanning trees of $G$). Instead of starting Mcuts-To-SpTrees on the row $(2,2,...,2)$ of length 50, we replaced $j$ random 2's by 1's  and started Mcuts-To-SpTrees on the resulting row $r$. For each $(1\le j\le 8)$ we launched three trials and recorded the following parameters for the one trial that produced the median number of spanning trees:
The cardinality\footnote{It sometimes happened that some of the $j$ many 1-bits matched the edges of a cycle. Then $r\cap ST(G)=\emptyset$; but this was not taken into account and another random $j$-set was chosen.} of $r\cap ST(G)$, the number of produced disjoint 012e-rows, and the CPU-time. 
The explanation for the times, in a nutshell, is as follows. The more 1-bits a row has\footnote{What about the 0-bits? Increasing the number of 0-bits increases (slowly) the risk that some mincut falls into $zeros(r)$. In this case we drop $r$ and turn to the next 012-row provided by {\tt BooleanConvert}. Hence 0's are irrelevant to make the case in Table 7.  }, the more likely is it that a mincut contains (the position of) the 1-bit, and so needs not be imposed., and so speeds up Mcuts-To-SpTrees.

\begin{tabular}{|c|c|c|c|  }
	 $\#$ of 1-bits (=j) & $\#$ of spanning trees & $\#$ of 012e-rows & time in sec     \\ 
	  &  &  &      \\ \hline
	 8 & 140 & 2 & 0.02    \\ \hline
	 7 & 1684 & 6 & 0.06      \\ \hline
	  6 & 16269 & 24 & 0.06      \\ \hline
	  5 & 74589 & 96 & 0.81      \\ \hline
	  4& 590874 & 408 & 4      \\ \hline
	  3 & 6'287'138 & 4320 & 39      \\ \hline
	  2 & 33'018'480 & 26640 & 241      \\ \hline
	  1 & 172'497'600 & 252000 & 2131      \\ \hline
	  \end{tabular}

{\sl Table 7. The fewer 2's are contained in a 012-row, the faster Mcuts-To-SpTrees runs on it.}

{\bf 6.4} In 6.2.1 we had the situation that the starter rows could not be produced in polynomial total time. Here comes a scenario where they can be produced quickly. Suppose we want to enumerate the family $\cal D$ of all spanning trees $T\su E$ of $G=(V,E)$ satisfying a list of conditions $A_1\rightsquigarrow B_1,...,A_t\rightsquigarrow B_t$. By definition this means that $T\in ST(G)$ only qualifies iff $A_i\cap T=\emptyset$ implies $B_i\cap T\neq \emptyset$ for all $1\le i\le t$. Intuitively: If all desirable edges in $A_i$ are absent in $T$ then at least one of the "safeguards" in $B_i$ must be present. (Real-life applications are easy to come by.)
Without proof, the family of all subsets $X\su E$ satisfying the conditions $A_i\rightsquigarrow B_i$ can be represented as a disjoint union\footnote{For instance, if $G=G_1$ and $t=1$ and $A_1\rightsquigarrow B_1$ is $\{5\}\rightsquigarrow\{1,7\}$ then $p=2$ with $r_1\uplus r_2=(e,2,2,2,{\bf 0},2,e)\uplus (2,2,2,2,{\bf 1},2,2)$.  We mention that $T=\{2,3,4,6\}$ happens to be the only spanning tree of $G_1$ not satisfying $\{5\}\rightsquigarrow\{1,7\}$; thus $\{5\}\cap T=\{1,7\}\cap T=\emptyset$.} $r_1\uplus\cdots\uplus r_p$ of 012e-rows in polynomial total time. In brief, that's because the conditions $A_i\rightsquigarrow B_i$ are dual to well-behaved  {\it implications} (aka pure Horn functions) $A_i\to B_i$ which by definition are satisfied by $T$ iff from  $A_i\su T$ follows $B_i\su T$.

\section{About edge-covers}

 What about freeing Mcuts-To-SpTrees from its (too) many mincuts altogether while maintaining the $e$-algorithm? Specifically, let $G(V,E)$ be a graph.  If $X\su E$ cuts all sets $star(v)$ (i.e. $X$ is incident with all vertices), then $X$ is called an {\it edge-cover}.
Feeding the $e$-algorithm with the merely $n$ sets $star(v)$ (instead of up to $ 2^{n-1}-1$ mincuts) renders the set $ECOV(G)$ of all edge-covers as a disjoint union of disjoint 012e-rows $r_i$. One checks that the spanning trees of $G$ are exactly those $(n-1)$-element edge-covers that are connected (equivalently: that are acyclic). The spanning trees can hence be enumerated  by scanning for each $r_i$ its $(n-1)$-element subsets, and picking those that happen to be connected. Call this method {\it Stars-To-SpTrees}. For the two random graphs in Table 8 it severly beats Mcuts-To-SpTrees but takes about double the time of Winter's algorithm. 

\begin{tabular}{|c|c|c|c|c|c|c|  }
  (n,m) & $|ST(G)|$ &  Winter & McutsToSpTrees & (n-1)-ecs & StarsToSpTrees  &  StarsToMecs   \\ 
  &  &  &  &  &  &  \\ \hline
 (15,25) & 175152 &23 sec& 155 sec & 746952  & 55 sec    & 3606 in 0.8 sec \\ \hline
 (15,27) & 708313 &93 sec& 1780 sec & 2'638'075  & 204 sec    & 4734 in 1.2 sec \\ \hline 
\end{tabular}

{\sl Table 8: Comparing three algorithms that are based on $e$-wildcards.}

At present  Stars-To-SpTrees sieves the trees one-by-one. For instance, the (15,27)-graph has 2'638'075 $(n-1)$-element edge-covers, among which 708313 turned out to be spanning trees. It is conceivable that some future enhancement achieves some kind of compression.

{\bf 7.1}  While $CONN(G)\su ECOV(G)$, there tend to be much fewer minimal edge-covers (=: {\it mecs}) than minimal connected sets (=spanning trees). Thus  the (15,27)-graph has 708313 spanning trees but only 4734 mecs. This suggests to persue minimal edge-covers as competitors to spanning trees.
 As opposed to spanning trees, mecs come in different cardinalities. At present our method, call it {\it Stars-To-Mecs} simply scans all row-minimal sets\footnote{ There were 6740 row-minimals in the (15,27)-graph, distributed among 545 final 012e-rows. Of these 304 were  very-good, 226 were good, and merely 15 were bad. At present, and much different from [W2], the  very-good rows $r$ can only be identified as such once all row-minimals in $r$ have been checked individually.} and sieves the ones that are minimal. Deciding minimality is straightforward, details will follow in some future work.
Notice that finding the 4734 mecs took 1.2 seconds, considerably faster than the 93 seconds it took Winter's method to get all spanning trees.

If a graph $G$ has proper matchings, then these are its minimum-cardinality mecs. Hence (see 3.2.2), upon running Stars-To-Mecs on $G$, all proper matchings show up in very-good rows, and so are easily detected (further improvements are possible).

In view of these preliminary findings, it is worthwile pondering in which of the countless applications of spanning trees the latter are  replaceable, with little harm (or even benefit), by the less numerous mecs.

\vspace{10cm}

{\bf References}

\begin{enumerate}

\item[{[CCCMP]}] M. Chakraborty, S. Chowdhury, J. Chakraborty, R. Mehera, RK Pal, Algorithms for generating all possible spanning trees of a simple undirected connected graph: an extensive review, Complex and Intelligent Systems 2018.
\item[{[C]}] C.J. Colbourn, The combinatorics of network reliability, Oxford University Press 1987.

\item[{[K]}] D. Knuth, The Art of Computer Programming, Volume 4A, Combinatorial Algorithms Part 1, Addison Wiley 2012.

\item[{[TSOA]}] S. Tsukiyama, I. Shirakawa, H. Ozaki, H. Ariyoshi, An algorithm to enumerate all cutsets of a graph in linear time per cutset, Journal of the ACM 27 (1980) 619-632.

\item[{[W1]}] M. Wild, Counting or producing all fixed cardinality transversals, Algorithmica 69 (2014) 117-129.

\item[{[W2]}] M. Wild, Compression with wildcards: All exact, or all minimal hitting sets, arXiv:2008.08996.

\item[{[W3]}] M. Wild, Compression with wildcards: All spanning trees, arXiv:2002.09707.

\item[{[Wi]}] P. Winter, An algorithm for the enumeration of spanning trees, BIT Numer. Math. 26 (1986) 44-62.

\end{enumerate}

\end{document}